\documentclass[10pt,letterpaper,english]{article}
\usepackage[T1]{fontenc}
\usepackage[latin1]{inputenc}
\usepackage{amsmath}
\usepackage{graphicx}
\usepackage{amssymb}

\makeatletter
\makeatletter%%%%%%%%%%%%%%%%%%%%%%%%%%%%%% User specified LaTeX commands.
\usepackage{amscd}

\usepackage{amsfonts}

\usepackage{babel}\makeatother

\makeatother

\usepackage{babel}

\begin{document}
\newtheorem{proposition}{Proposition}[section] \newtheorem{definition}{Definition}[section]
\newtheorem{corollary}{Corollary}[section] \newtheorem{lemma}{Lemma}[section]
\newtheorem{theorem}{Theorem}[section] \newtheorem{example}{Example}[section]
%$\left(\kern-3.2truept\begin{array}{c}a\\b\end{array}\kern-3.2truept\right)$

\title{\textbf{Dynamic optimization and its relation to \\ classical and quantum constrained systems.}}

\author{Mauricio Contreras, Rely Pellicer and Marcelo Villena.%
\thanks{Faculty of Engineering \& Sciences, Universidad Adolfo Ibañez, Chile.%
}}

\maketitle
\noindent
%\underline{\ \ \ \ \ \ \ \ \ \ \ \ \ \ \ \ \ \ \ \ \ \ \ \ \ \ \ \ \  \ \ \ \ \ \ \ \ \ \ \ \ \ \ \ \ \ \ \ \ \ \ \ \ \ \ \ \ \ \ \ \ \ \ \ \ \ \ \ \ \ \ \ \ \ \ \ \ \ \ \ \ \ \ \ \ %\ \ \ } \\ \\
%Published in Physica A: Statistical Mechanics and its Applications\\
%Volume 389, 2010, Pages 3552-3564 \\
%http://www.sciencedirect.com/science/article/pii/S0378437110003572\\
%DOI information: doi:10.1016/j.physa.2010.04.019\\
%\underline{\ \ \ \ \ \ \ \ \ \ \ \ \ \ \ \ \ \ \ \ \ \ \ \ \ \ \ \ \  \ \ \ \ \ \ \ \ \ \ \ \ \ \ \ \ \ \ \ \ \ \ \ \ \ \ \ \ \ \ \ \ \ \ \ \ \ \ \ \ \ \ \ \ \ \ \ \ \ \ \ \ \ \ \ \ %\ \ \ } \\ \\
We study the structure of a simple dynamic optimization problem consisting of one state and one control variable, from a physicist's point of view. By using an analogy to a physical model, we study this system in the classical and quantum frameworks. Classically, the dynamic optimization problem is equivalent to a classical mechanics constrained system, so we must use the Dirac method to analyze it in a correct way. We find that there are two second-class constraints in the model: one fix the momenta associated with the control variables, and the other is a reminder of the optimal control law. The dynamic evolution of this constrained system is given by the Dirac's bracket of the canonical variables with the Hamiltonian. This dynamic results to be identical to the unconstrained one given by the Pontryagin equations, which are the correct classical equations of motion for our physical optimization problem. In the same Pontryagin scheme, by imposing a closed-loop $\lambda$-strategy, the optimality condition for the action gives a consistency relation, which is associated to the Hamilton-Jacobi-Bellman equation of the dynamic programming method. A similar result is achieved by quantizing the classical model. By setting the wave function  $\Psi(x,t) = e^{iS(x,t)}$ in the quantum Schr\"{o}dinger equation, a non-linear partial equation is obtained for the $S$ function. For the right-hand side quantization, this is the Hamilton-Jacobi-Bellman equation, when $S(x,t)$ is identified with the optimal value function. Thus, the Hamilton-Jacobi-Bellman equation in Bellman's maximum principle, can be interpreted as the quantum approach of the optimization problem.

\section{Introduction}

We have recently witnessed the increasing application of ideas from physics to finance and economics, such as path integral techniques applied to study the Black-Scholes model in its different forms \cite{Cbennatirosaclot}, \cite{Otto}, \cite{lemmens1}, \cite{baaquie}, \cite{baaquie2}, \cite{Linetsky}, \cite{Dash}. Some developments have also been used to try to understand the Black-Scholes equation as a quantum mechanical Schr\"{o}dinger equation \cite{Haven1}, \cite{Haven2}, \cite{semiclass}. In the last few years, constrained systems techniques, through Dirac's method \cite{dirac1}, \cite{dirac2} have been used to explain some features of stochastic
volatility models \cite{contrerashojman}, \cite{contreras} and the multi-asset Black-Scholes equation \cite{contrerasbustamante}. In this paper, we apply these same constrained methods to understand (from a physical point of view) a particular issue: the dynamic optimization problem. \\  \\
We start by analyzing the dynamic optimization problem for a single-state variable $x$ and a control variable $u$. By identifying
the state variable $x$ as the coordinate of a physical particle and the Lagrange multiplier $\lambda$ as its canonical momentum $p_{x}$,
we can map the theory in the Hamiltonian phase space. Here, the model presents constraints thus, it is necessary (to study the system
in a correct way) to use Dirac's method of constrained systems. The application of this method implies that constraints are of
second-class character according to Dirac's classification. Thus, the dynamic optimization problem can be seen as a second-class physically constrained system. \\
We also analyze the role of open-loop and closed-loop strategies in the context of Pontryagin's framework. We explicitly show that the only consistent strategies that permit the Pontryagin equations to be obtained correctly from the optimization of cost functionals are open-loop $\lambda$-strategies ($\lambda=\lambda(t)$). For closed loop $\lambda$-strategies ($\lambda=\lambda(x,t)$), the optimization of the cost functional gives a consistency relation which is related to the Hamilton-Jacobi-Bellman equation. \\ \\
After that, we explore the quantum side of this classically constrained system. By quantizing it according to the standard rules of quantum mechanics and imposing the constraints as operator equations over the wave function, we arrive at a set of partial differential equations for the wave function. After defining the wave function as $\Psi(x,t)=e^{iS(x,t)}$, these equations map into some partial differential equations for the $S$ function. For right-hand side quantization order, these equations give origin to the Hamilton-Jacobi-Bellman equation of the dynamic programming theory.  Thus, Bellman's maximum principle can be considered as the quantum view of the optimization problem. \\ \\
To make this paper self-contained for non-physicist readers coming from the optimization field, we start with a brief digression on classical and quantum physics in section II.

\section{Quantum and classical mechanics}
\subsection{Hamiltonian quantum and classical mechanics}
In physics, quantum-dynamic behavior is defined by the Hamiltonian operator. For the simple case of a nonrelativistic one-dimensional particle subjected to external potential $U(x)$, the Hamiltonian operator reads
\begin{equation*}
	\check{H} = \frac{1}{2m} {\check{P}_x}^2 + U(x) = -\frac{\hbar^2}{2m} \frac{\partial^2}{\partial x^2} +U(x),
\end{equation*}
where $\check{P}_x =- i \hbar \frac{\partial }{\partial x}$ is the momentum operator. The wave function at time $t$ (given that the wave function at $t=0$ is $\Psi_0$) is thus
\begin{equation*} \label{solutionSE}
	\Psi(x,t) = e^{-\frac{i}{\hbar}\check{H} t} \Psi_0(x),
\end{equation*}
which can be written as a convolution according to
\begin{equation*}
	\Psi(x,t) = \int K(x,t \vert x' 0) \Psi_0(x') dx',
\end{equation*}
where the propagator $K$ admits the Hamiltonian Feynman path integral representation:
\begin{equation} \label{propHami}%1
	K(x,t \vert x' 0) = \int {\cal{D}} x \frac{{\cal{D}} p_x}{2\pi} \exp(\frac{i}{\hbar} A[x, p_x] ).
\end{equation}
The symbol $ \int {\cal{D}} x \frac{{\cal{D}} p_x}{2\pi}$ denotes the sum over the set of all trajectories that start at $(x_0, p_{x}^0)$ for $t=0$ and end at $(x, p_{x})$ at time $t$ in the phase space. Thus, all trajectories are needed to evaluate the quantum propagator. In (\ref{propHami}), $A[x, p_x]$ is the classical Hamiltonian action functional, given by
\begin{equation*} \label{CAusual}
	A[x, p_x] = \int p_x \dot{x} - H(x,p_x)  \ dt,
\end{equation*}
where $H$ is the Hamiltonian function
\begin{equation} \label{classhami1}%2
	H(x, p_x)= \frac{1}{2m} p_x^2 + U(x).
\end{equation}
From the classical dynamics (Newtonian equations) of the quantum system, we understand the particular trajectory in the phase space $(x,p_x)$ that gives an extreme to the Hamiltonian functional $A[x, p_x]$: the path for which the variation of the action is zero:
\begin{equation*} \label{DeltaSusual}
	\delta A[x, p_x] = A[x+\delta x,p_x+\delta p_x] - A[x,p_x] = 0.
\end{equation*}
We can show that this classical trajectory necessarily satisfies the Hamiltonian equations of motion:
\begin{equation} \label{hamiltoneqxdot}%3
	\dot{x} = \frac{\partial H(x,p_x)}{\partial p_x},
\end{equation}
\begin{equation*} \label{hamiltonepxdot}
	\dot{p}_x = -\frac{\partial H(x,p_x)}{\partial x}.
\end{equation*}
These Hamiltonian equations can be written with Poisson brackets $\{ \ , \}$ as
\begin{equation*} \label{PBequation1}
	\dot{x} = \{ x, H \},\quad \dot{p}_x = \{ p_x, H \},
\end{equation*}
where
\begin{equation*} \label{PoissonBracket}
	\{ F, G \} = \frac{\partial F}{\partial x} \frac{\partial G}{\partial p_x} - \frac{\partial G}{\partial x} \frac{\partial F}{\partial p_x}
\end{equation*}
for any functions $F=F(x,p_x)$ and $G=G(x,p_x)$ defined over the phase space $(x, p_x)$. \\ \\
The dynamic evolution of a function $F=F(x,p_x)$ defined over the phase space is given by the classical counterpart of the Heisenberg equation:
\begin{equation*}
	\dot{F} = \{ F, H \} = \frac{\partial F}{\partial x} \frac{\partial H}{\partial p_x} - \frac{\partial H}{\partial x} \frac{\partial F}{\partial p_x}.
\end{equation*}
Using the Hamiltonian equations, this is
\begin{equation*}
	\dot{F} = \{ F, H \} = \frac{\partial F}{\partial x} \dot{x} + \frac{\partial F}{\partial p_x} \dot{p}_x.
\end{equation*}
Note that the Hamiltonian equations of motions are first-order differential equations; for example, for Hamiltonian (\ref{classhami1}), these equations are read as
\begin{equation}\label{xpunto}
	\dot{x} = \frac{\partial H(x,p_x)}{\partial p_x} = \frac{p_x}{m},
\end{equation}
\begin{equation}\label{ppunto}
	\dot{p}_x = -\frac{\partial H(x,p_x)}{\partial x} = \frac{\partial U}{\partial x}.
\end{equation}
Taking the time derivative of equation (\ref{xpunto}) and using (\ref{ppunto}), one can obtain Newton's equation of motion for a conservative system:
\begin{equation} \label{Newton equation}%4
	-\frac{\partial U}{\partial x}  =  m \frac{d^2 x}{dt^2}.
\end{equation}

\subsection{Lagrangian quantum and classical mechanics}
By integrating equation (\ref{propHami}) over the momentum variables, we arrive at the Feynman path integral representation for the quantum propagator:
\begin{equation} \label{propLagrangian}%5
	K(x,t \vert x' 0) = N \int {\cal{D}} x \exp(\frac{i}{\hbar} A[x, \dot{x}] ).
\end{equation}
Here, $N$ is a normalization constant coming from the momentum integration, and $A[x, \dot{x}]$ is the Lagrangian action functional
\begin{equation} \label{Lagrangianaction}%6
	A[x, \dot{x}] = \int L(x, \dot{x}) \ dt,
\end{equation}
where $L$ is the Lagrangian function defined by
\begin{equation} \label{Lagrangian}%7
	L[x, \dot{x}] = p_x \dot{x} - H(x,p_x)
\end{equation}
and where $p_x$ has been solved in terms of the velocity $\dot{x}$ by means of equation (\ref{hamiltoneqxdot}). \\ \\
For example, in the case of the Hamiltonian (\ref{classhami1}), equation (\ref{hamiltoneqxdot}) gives the relation
\begin{equation*}
	\dot{x} = \{ x, H \} = \frac{\partial H(x,p_x)}{\partial p_x} = \frac{1}{m} p_x;
\end{equation*}
that is,
\begin{equation} \label{momentumusual}%8
	p_x = m \dot{x},
\end{equation}
which, in this case, corresponds to the usual definition of momentum. Substituting equation (\ref{momentumusual}) in the Lagrangian (\ref{Lagrangian}) gives
\begin{equation} \label{Lagrangianusual}%9
	L(x, \dot{x}) = p_x \dot{x} - H(x,p_x) =  \frac{1}{2} m \dot{x}^2 - U(x),
\end{equation}
which is the common expression of the Lagrangian as the difference of the kinetic and potential energies. \\ \\
Classical mechanics, in the Lagrangian formalism, is linked to the particular path $x =x(t)$ that gives an extreme to the Lagrangian action functional (\ref{Lagrangianaction}):
\begin{equation*}
	\delta A[x, \dot{x}] = A[x+\delta x,\dot{x}+\delta \dot{x}] - A[x,\dot{x}] = 0.
\end{equation*}
This extremal condition implies that the classical path must satisfy the Euler-Lagrange equation of motion
\begin{equation*} \label{lagrangeeqution}
	\frac{\partial L}{\partial x} - \frac{d}{dt} \ \big( \frac{\partial L}{\partial \dot{x}}  \big)  =  0.
\end{equation*}
Note that for the Lagrangian (\ref{Lagrangianusual}), we have
\begin{equation} \label{momentum definition1}%10
	\frac{\partial L}{\partial \dot{x}}   =  m \dot{x} , \  \   \  \frac{\partial L}{\partial x}   = - \frac{\partial U}{\partial x},
\end{equation}
so the second-order Euler-Lagrange differential equation again gives the Newtonian equation of motion for a conservative system (\ref{Newton equation}). From equation (\ref{momentum definition1}), we see that
\begin{equation*}
	\frac{\partial L}{\partial \dot{x}}   =  m \dot{x} = p_x.
\end{equation*}
That is, $\frac{\partial L}{\partial \dot{x}}$ is the momentum of the  one-dimensional particle.  For a general Lagrangian $L(x, \dot{x})$, the relation
\begin{equation} \label{momentum definition}%11
	p_x = \frac{\partial L}{\partial \dot{x}}
\end{equation}
is a definition of the linear momentum of the system. In fact, (\ref{momentum definition}) is the Legendre transformation between the variables $(\dot{x}, p_x)$, communicating the Hamiltonian and Lagrangian classical approaches. In Hamiltonian theory, the analogous role of the Legendre transformation is given by the first Hamiltonian equation (\ref{hamiltoneqxdot}). \\ \\
When equation (\ref{momentum definition}), in the Lagrangian context, can be written in terms of the velocity $\dot{x}$ or, in the Hamiltonian context,  (\ref{hamiltoneqxdot}) can be solved for the momentum $p_x$, both approaches are equivalent. Thus, equations (\ref{momentum definition}) and (\ref{hamiltoneqxdot}) are a bridge between the Hamiltonian and Lagrangian theories, and that bridge is open when the momentum can be solved  in terms of the velocity or vice-versa. \\ \\
However, there are many situations for which this is not possible, and those are the most important ones, as we shall see later. A typical situation occurs, for instance, when the Lagrangian is a  linear function of the velocities, as in
\begin{equation*}
	L(x, \dot{x}) = a \dot{x} + b.
\end{equation*}
In this case, the momentum is
\begin{equation} \label{constraint1}%12
	p_x = \frac{\partial L}{\partial \dot{x}} = a,
\end{equation}
from which the momentum cannot be solved in terms of the velocity $\dot{x}$. Equation (\ref{constraint1}) is a constraint on the momentum variable in the phase space; it fixes $p_x$ to take the value $a$ at all times. \\ \\
In physics, the class of systems for which the Legendre bridges (\ref{momentum definition}) or (\ref{hamiltoneqxdot}) are closed (that is, when the momentum cannot be solved  in terms of the velocity or vice-versa) are called constrained systems. In this case, the classical Lagrangian and Hamiltonian theories are not equivalent in general. In a quantum setting, the Lagrangian and Hamiltonian propagators for a constrained system, (\ref{propLagrangian}) and (\ref{propHami}) are not equal because the naive integration measure in the Lagrangian/Hamiltonian Feynman path integral is incorrect due to the presence of constraints. \\ \\
Paul Maurice Dirac developed a strategy that allows for the study of general constrained systems in the phase space. This method is now called Dirac's method \cite{dirac1}, \cite{dirac2}. In section IV, we will apply this method to study the Pontryagin equations regarding the motion of a second-class constrained physical system.

\section{Dynamic optimization problem}
\subsection{The Pontryagin approach}
Consider an optimal control problem that is commonly used in financial applications (see, for example, \cite{Kamien}). We want to optimize the cost functional
\begin{equation*}
	A[x, u]  = \int_{t_0}^{t_1} F(x,u,t) \ dt,
\end{equation*}
where $x$ represents a state variable (for example, the production of a certain article) and $u$ is a control variable (such as the marketing cost). The state variable must satisfy the market dynamic
\begin{equation} \label{xdotf}%13
	\dot{x} = f(x,u,t).
\end{equation}
The problem is to determine how to obtain the production trajectory $x=x(t)$ and the control path $u=u(t)$ to optimize the cost functional. To get the solution, we apply the Lagrange multipliers method, so we consider the improved functional $A$ on the extended configuration space $(x,u,\lambda)$, which is defined by
\begin{equation} \label{A}%14
	A[x,u, \lambda] = \int_{t_0}^{t_1} F(x,u,t) - \lambda(\dot{x} - f(x,u,t)) \ dt.
\end{equation}
To obtain the solution for our problem, we can interpret the integrand of (\ref{A}) as the Lagrangian
\begin{equation} \label{Lagrangiancontopt1}
	L(x,u, \lambda, \dot{x}, \dot{u},\dot{\lambda}) =  F(x,u,t) - \lambda(\dot{x} - f(x,u,t)).
\end{equation}
The extremal curves then satisfy the Euler-Lagrange equations:
\begin{equation*}
	\frac{\partial L}{\partial \lambda} - \frac{d}{dt} \ \big( \frac{\partial L}{\partial \dot{\lambda}}  \big)  =  0
\end{equation*}
\begin{equation*}
	\frac{\partial L}{\partial x} - \frac{d}{dt} \ \big( \frac{\partial L}{\partial \dot{x}}  \big)  =  0
\end{equation*}
\begin{equation*}
	\frac{\partial L}{\partial u} - \frac{d}{dt} \ \big( \frac{\partial L}{\partial \dot{u}}  \big)  =  0.
\end{equation*}
These are also written as, respectively,
\begin{equation*}
	\dot{x} - f(x,u,t)  =  0
\end{equation*}
\begin{equation*}
	\frac{\partial}{\partial x} \big( F +  \lambda f \big) + \dot{\lambda} =  0
\end{equation*}
\begin{equation*}
	\frac{\partial}{\partial u} \big( F +  \lambda f \big) =  0,
\end{equation*}
or as
\begin{equation} \label{pontryx}%15
	\dot{x} = \frac{\partial H}{\partial \lambda}
\end{equation}
\begin{equation} \label{pontrylambda}%16
	\dot{\lambda} = - \frac{\partial H}{\partial x}
\end{equation}
\begin{equation} \label{pontryu}%17
	\frac{\partial H}{\partial u} =  0,
\end{equation}
with $H$ defined by
\begin{equation*}
	H = H(x, u, \lambda) = F(x,u,t) + \lambda f(x,u,t).
\end{equation*}
Equations (\ref{pontryx}), (\ref{pontrylambda}) and (\ref{pontryu}) are the well-known Pontryagin equations; they are obtained through the Pontryagin maximum principle. Note that the first two equations, (\ref{pontryx}) and (\ref{pontrylambda}), are just the Hamilton equations of motion if we interpret the Lagrange multiplier $\lambda$ as the canonical momentum $p_x$ associated to the state variable $x$. Thus, by identifying $\lambda$ with $p_x$, we can rewrite the Lagrangian (\ref{Lagrangiancontopt1}) as
\begin{equation} \label{Lagrangiancontopt2}
	L(x,u, \dot{x}, \dot{u}) = - p_x \dot{x} + ( F(x,u,t) + p_x f(x,u,t)).
\end{equation}
The above Lagrangian is minus the Lagrangian of a physical particle in classical mechanics, so dynamic optimization is completely equivalent to a classical physical system at both the Lagrangian and Hamiltonian levels.

\subsection{Open-loop and closed-loop strategies}
The action (\ref{A}) can be written in a compact form as
\begin{equation} \label{A2}
	A[x,u, \lambda] = \int_{t_0}^{t_1}   - \lambda \dot{x} +  H(x,u,\lambda,t) \ dt.
\end{equation}
Strictly, the Pontryagin equations must be obtained by optimizing action (\ref{A2}) with respect to its three variables $x,u,\lambda$; that is, we consider
\begin{eqnarray*}
	A[x+\delta x,u+\delta u, \lambda+ \delta \lambda] = & \\
	\int_{t_0}^{t_1} -[\lambda + \delta \lambda] [ \dot{x} + \dot{\delta x} ] +  H &(x+\delta x,u+\delta u,\lambda  + \delta \lambda,t) \ dt,
\end{eqnarray*}
where $\delta x, \delta u$ and $\delta \lambda$ are the corresponding functional  variations of the initial variables. Now, expanding the Hamiltonian in a Taylor series and keeping the first-order terms only, we have
\begin{equation*}
	\delta A = \int_{t_0}^{t_1}  \Bigl[( \frac{\partial H}{\partial \lambda} - \dot{x}  ) \delta \lambda - \lambda \dot{\delta x} + \frac{\partial H}{\partial x } \delta x + \frac{\partial H}{\partial u} \delta u  \Bigr] dt.
\end{equation*}
Finally, integrating by parts, we get
\begin{eqnarray} \label{A3}
	\nonumber \delta A & = & \displaystyle\int_{t_0}^{t_1} \Bigl[ ( \frac{\partial H}{\partial \lambda} - \dot{x}  ) \delta \lambda + (\frac{\partial H}{\partial x } + \dot{\lambda} ) \delta x + \frac{\partial H}{\partial u} \delta u \Bigr] dt \\
	& - & \lambda(t_1) \delta x(t_1).
\end{eqnarray}
To maximize the action, all the first-order terms in $\delta x, \delta u$ and $\delta \lambda$ must vanish. If the variables $x, u$ and $\lambda$ are independent, then $\delta x, \delta u$ and $\delta \lambda$ are linearly independent, so we obtain the Pontryagin equations and the transversality conditions $\lambda(t_1) = 0$ from equation (\ref{A3}). \\ \\
Now, it is well-known that two classes of control strategies exist: open-loop strategies that depend only on time, such as
\begin{equation*}
	u=u(t),
\end{equation*}
and closed-loop strategies that depend on the state variable $x$ and on time (see, for example, \cite{erikson1}):
\begin{equation*}
	u=u(x,t).
\end{equation*}
For an open-loop strategy, the variational principle for (\ref{A2}) is well-defined because the variables
\begin{equation*}
	x=x(t), u=u(t)\;\mbox{and}\; \lambda=\lambda(t)
\end{equation*}
remain independent and because the Pontryagin equations can be obtained correctly from (\ref{A3}). What happens, however, with closed-loop strategies? In this case, due to the relationships between the variables in $u=u(x,t)$, the functional variations are related by
\begin{equation*}
	\delta u = \frac{\partial u}{\partial x}  \delta x.
\end{equation*}
Substituting this into (\ref{A3}), we obtain
\begin{eqnarray*}
	\nonumber \delta A & = & \displaystyle\int_{t_0}^{t_1} \Bigl[ ( \frac{\partial H}{\partial \lambda} - \dot{x}  ) \delta \lambda + (\frac{\partial H}{\partial x } +  \frac{\partial H}{\partial u} \frac{\partial u}{\partial x} + \dot{\lambda} ) \delta x \Bigr] dt \\
	& - & \lambda(t_1) \delta x(t_1).
\end{eqnarray*}
If $\lambda$ and $x$ remain independent, we arrive at
\begin{equation*}
	\frac{\partial H}{\partial \lambda} - \dot{x} = 0\;,\qquad\quad\frac{\partial H}{\partial x } +  \frac{\partial H}{\partial u} \frac{\partial u}{\partial x} + \dot{\lambda} = 0,
\end{equation*}
and the transversality condition,  but the equation that gives the optimal condition for the control is lost. Then,  if $u$ is not known as a
function of $x$ from the beginning, we have three unknowns $x, u$ and $\lambda$ but only two equations of motion. \\ \\
Now, the control equation $\frac{\partial H}{\partial u} = 0$ is in fact the following algebraic equation for $u$
\begin{equation} \label{pontryuv2}
	\frac{\partial F(x,u, t)}{\partial u} + \lambda  \ \frac{\partial f(x,u,t)}{\partial u} = 0;
\end{equation}
from this, the control $u$ can be obtained as a function of $x$ and $\lambda$:
\begin{equation}\label{uxl}
	u = u^{*}(x, \lambda,t),
\end{equation}
and the same optimization problem implies that $u$ is a closed-loop strategy! How can this be consistent with the variational problem in which $x$, $u$ and $\lambda$ are independent variables? \\ \\
To see this, consider a general closed-loop strategy of the form $u = u(x, \lambda,t)$, then we have
\begin{equation*}
	\delta u = \frac{\partial u}{\partial x} \delta x + \frac{\partial u}{\partial \lambda} \delta \lambda.
\end{equation*}
After substituting this in (\ref{A3}), $\delta A $ equals
\small\begin{eqnarray} \label{A3v2}
	\nonumber \displaystyle\int_{t_0}^{t_1}\Bigl[(\displaystyle\frac{\partial H}{\partial \lambda} & + \displaystyle\frac{\partial H}{\partial u} \frac{\partial u}{\partial \lambda}  - \dot{x}  ) \delta \lambda + (\frac{\partial H}{\partial x } +  \frac{\partial H}{\partial u} \frac{\partial u}{\partial x} + \dot{\lambda} ) \delta x  \Bigr]dt \\
	- \lambda(t_1)\delta x(t_1), &
\end{eqnarray}
\normalsize thus, if $x$ and $\lambda$ are independent variables, we have the equations of motions from the variational principle:
\begin{equation} \label{ponl2}
	\frac{\partial H}{\partial \lambda} + \frac{\partial H}{\partial u} \frac{\partial u}{\partial \lambda}  - \dot{x} = 0,
\end{equation}
\begin{equation} \label{ponx2}
	\frac{\partial H}{\partial x } +  \frac{\partial H}{\partial u} \frac{\partial u}{\partial x} + \dot{\lambda} = 0.
\end{equation}
Obviously, these equations, for an arbitrary closed-loop strategy $u=u(x, \lambda)$, differ from the Pontryagin open loop equations. If we choose for $u$ the special strategy $u^{*}$ that is the solution of (\ref{pontryuv2}), however (\ref{ponl2}) and (\ref{ponx2}) reduce to Pontryagin equations and the system is consistent. On the other hand, we can think of $x,u$ and $\lambda$ as independent or of $u$ as the closed-loop strategy that is the solution of (\ref{pontryu}) or (\ref{pontryuv2}). In a sense, the special closed-loop strategy $u=u^{*}$ is completely equivalent to an open-loop strategy. \\ \\
Note that for an arbitrary close-loop strategy, equations (\ref{ponl2}) and (\ref{ponx2}) mean that the action is optimized. But, these extremes are not necessarily global. Condition (\ref{pontryu}) gives the global extreme for the action $A$.\\ \\
To end this section, suppose that $\lambda$ and $x$ are not independent and they are related by $\lambda = \lambda(x,t)$; then, the variation of $\lambda$ is
\begin{equation*}
	\delta \lambda = \frac{\partial \lambda}{\partial x} \ \delta x.
\end{equation*}
Substituting this into (\ref{A3v2}) and using $u=u^{*}$, $\lambda(t_1) = 0$, and $\dot{\lambda} = \displaystyle\frac{\partial \lambda }{\partial x} \dot{x} + \frac{\partial \lambda }{\partial t}$, we get
\begin{equation} \label{A3v3}
	\delta A = \int_{t_0}^{t_1}  \Bigl[ ( \frac{\partial H}{\partial \lambda} \frac{\partial \lambda }{\partial x} - \dot{x} \frac{\partial \lambda }{\partial x} ) + (\frac{\partial H}{\partial x } + \frac{\partial \lambda }{\partial x} \dot{x} + \frac{\partial \lambda }{\partial t} )  \Bigr] \delta x  \ \ dt,
\end{equation}
but, as  $H=F(x,u^{*},t)+\lambda f(x,u^{*},t)$, so too is
\begin{equation*}
	\frac{\partial H}{\partial \lambda} = f(x,u^{*},t,)
\end{equation*}
\begin{equation*}
	\frac{\partial H}{\partial x } = \frac{\partial F(x,u^{*},t)}{\partial x } + \lambda \frac{\partial f(x,u^{*},t)}{\partial x }.
\end{equation*}
Thus, (\ref{A3v3}) gives finally
\begin{equation*} \label{A3v4}
	\delta A =  \int_{t_0}^{t_1}  \Bigl[ (f \frac{\partial \lambda }{\partial x} + \frac{\partial F}{\partial x } +  \lambda \frac{\partial  f }{\partial x}) + \frac{\partial \lambda }{\partial t}   \Bigr] \delta x  \ \ dt
\end{equation*}
or
\begin{equation*} \label{A3v5}
	\delta A =  \int_{t_0}^{t_1}  \Bigl[ \frac{d H^* }{d x} + \frac{\partial \lambda }{\partial t}   \Bigr] \delta x  \ \ dt.
\end{equation*}
where
\begin{equation} \label{A3v66}
	H^*= H^*(x,t) = H(x, u^*(x, \lambda(x,t), t), \lambda(x,t), t)
\end{equation}
is the reduced Hamiltonian in terms of $x$. In this way, the optimization of the action $A$, implies that the closed-loop $\lambda = \lambda(x,t)$ strategy must satisfy the following optimal consistency condition
\begin{equation} \label{A3v6}
	\frac{ d H^*(x,t) }{ d x} + \frac{\partial \lambda }{\partial t}  = 0. \end{equation}
We will see in next section that (\ref{A3v6}) is closely related to the Hamilton-Jacobi-Bellman equation. Let $\lambda^{*}(x,t)$ be a solution of (\ref{A3v6}); then, the optimal state variable $x(t)$ can be obtained from (\ref{xdotf}), according to
\begin{equation} \label{xdotf2}
	\dot{x} = f(x,u^{*}(x,\lambda^{*}(x,t), t), t).
\end{equation}
Note that (\ref{xdotf2}) can be viewed as the Pontryagin equation for $x(t)$, in the sense that $u^{*}$ and $\lambda^{*}$  are chosen in a way that they are optimal closed-loop strategies, that is, these strategies maximizes or minimizes the action.
We make the following points to conclude this section: \\ \\
1) The optimal closed-loop control strategies $u^{*}=u^{*}(x,\lambda)$ that satisfy $ \frac{\partial H}{\partial u} = 0 $ can be seen as open-loop strategies because they give the same equations of motion as in the open-loop case. We can say that they are inert because they do not change the open-loop dynamics. \\ \\
2) When the Lagrange multiplier $\lambda$ and the state variable $x$ are independent (we say that we have an open-loop $\lambda$-strategy), then $\lambda = \lambda(t)$ and the solution of the optimization problem is given by the Pontryagin equations.\\ \\
3) If there is a relationship between $\lambda$ and $x$ as $\lambda = \lambda(x,t)$ (we say that we have a closed-loop $\lambda$-strategy), this strategy will be optimal only if it satisfy the consistency condition (\ref{A3v6}). The solution of the optimization problem is given by the Pontryagin equation (\ref{xdotf2}), where $\lambda^{*}$ satisfy (\ref{A3v6}).

\subsection{The Bellman approach}

A second approach to the optimization problem comes from dynamic programming theory, and was developed by Richard Bellman \cite{Bellman}.  In this case, the fundamental variable is the optimal value of the action defined by
\begin{equation}
	J(x_0, t_0) = \max_u \Bigg( \int_{t_0}^{t} F(x, u, t) \ dt  \Bigg) ,
\end{equation}
subject to (\ref{xdotf}), with initial condition $x(t_0) = x_0$. \\ \\
The optimality principle of Bellman implies that $J(x,t)$ satisfies the Hamilton-Jacobi-Bellman equation (see \cite{Kamien}):
\begin{equation} \label{HJB}
	\max_u \Bigl( F(x,u,t) + \frac{\partial J(x,t) }{\partial x}  f(x,u,t) \Bigr) = - \frac{\partial J(x,t) }{\partial t} .
\end{equation}
The left-hand side of equation (\ref{HJB}) is just the maximization of the Hamiltonian (\ref{pontryuv2}) with respect to the control variable $u$, where the Lagrangian multiplier $\lambda$ of the Pontryagin approach must be identified with $\frac{\partial J(x,t)}{\partial x}$. Thus, the Bellman theory can be seen from the Pontriagyn perspective as a closed-loop $\lambda$-strategy $\lambda(x,t) = \frac{\partial J(x,t)}{\partial x}$. \\ \\
By maximizing and solving for the optimal control variable in the left-hand side of (\ref{HJB}) as $u^*= u^*(x,t)=u^*(x,\lambda(x,t),t)=u^*(x,\frac{\partial J(x,t) }{\partial x},t)$, the Hamilton-Jacobi-Bellman equation is \\
\begin{equation} \label{HJBv3}
	F(x,u^*,t) + \frac{\partial J(x,t) }{\partial x}  f(x,u^*,t) = - \frac{\partial J(x,t) }{\partial t} .
\end{equation}
If we differentiate in (\ref{HJBv3}) with respect to $x$, we get
\begin{small}$$\begin{array}{l}
	\displaystyle\frac{\partial F(x,u^*,t) }{\partial x} + \displaystyle\frac{\partial^2 J(x,t) }{\partial x^2}  f(x,u^*,t) + \frac{\partial J(x,t) }{\partial x}  \frac{\partial f(x,u^*,t) }{\partial x} \\
	\\
	+\Bigg(\displaystyle\frac{\partial F(x,u^*,t) }{\partial u^*} + \displaystyle\frac{\partial J(x,t) }{\partial x} \frac{\partial f(x,u^*,t) }{\partial u^*} \Bigg) \frac{d u^*(x,t)}{dx}    \\ \\
	= - \displaystyle\frac{\partial^2 J(x,t) }{\partial x \partial t } .
	\end{array}$$\end{small}
Using the fact that $u^*$ is optimal and replacing $\lambda(x,t)=\frac{\partial J(x,t)}{\partial x}$ we obtain
$$\begin{array}{l}
\displaystyle\frac{\partial F(x,u^*,t) }{\partial x} + \frac{\partial \lambda(x,t) }{\partial x}  f(x,u^*,t) + \lambda \frac{\partial f(x,u^*,t) }{\partial x}   \\ \\
= - \displaystyle\frac{\partial \lambda(x,t) }{\partial t} ,
\end{array}$$
or
\begin{equation*}
	\frac{d H^*(x,t)}{dx} + \frac{\partial \lambda(x,t) }{\partial t}  = 0 .
\end{equation*}
This equation is identical to (\ref{A3v6}). Thus, this optimal consistency condition for the closed-loop $\lambda^*$-strategy is nothing more but the derivative of the Hamilton-Jacobi-Bellman equation. Then, equation (\ref{A3v6}) can be written, according to (\ref{HJBv3}), as
\begin{equation}
	\frac{d}{dx} \Bigg( F(x,u^*,t) + \frac{\partial J(x,t) }{\partial x}  f(x,u^*,t) + \frac{\partial J(x,t) }{\partial t}  \Bigg)  = 0 .
\end{equation}
Integrating in the above equation, gives finally
\begin{equation*}
	F(x,u^*,t) + \frac{\partial J(x,t) }{\partial x}  f(x,u^*,t) + \frac{\partial J(x,t) }{\partial t} = g(t),
\end{equation*}
where $g(t)$ is an arbitrary, time-dependent function. \\ \\
Thus, for an optimal closed-loop $\lambda^*$-strategy, the Pontryagin optimal scheme gives an non-homogeneous Hamilton-Jacobi-Bellman equation. The Bellman maximum principle instead, gives an homogeneous Hamilton-Jacobi-Bellman equation. We conclude this section by saying that for a closed-loop $\lambda^*$-strategy, both (Pontryagin's and Bellman's) optimal approaches, are equivalent modulus an arbitrary time-dependent function.

\section{The optimization problem as a classically constrained system}
From a structural point of view, the optimization problem is then characterized completely by the Lagrangian multiplier $\lambda$. For a open-loop $\lambda^*$-strategy, the optimization of the action (\ref{A}) gives a system of coupled ordinary differential equations: the Pontryagin equations (\ref{pontryx}) and (\ref{pontrylambda}) for both, open or closed-loop optimal $u^*$-strategies. For a closed-loop $\lambda^*$-strategy instead, the optimization of the action (\ref{A}) gives a partial differential equation for $\lambda$: the consistency relation (\ref{A3v6}), which is equivalent (modulus an arbitrary time-dependent function) to the Hamilton-Jacobi-Bellman equation of the dynamic programming theory. \\ \\
From a physical point of view, the Pontryagin equations (\ref{pontryx}) and (\ref{pontrylambda}) can be seen as the Hamilton equations of a classical mechanical system, if we identify the Lagrangian multiplier $\lambda$ with the canonical momentum $p_x$ of the $x$ variable. The Hamilton-Jacobi-Bellman equation (\ref{HJBv3}) instead, looks like a Schr\"{o}dinger equation for the wave function $J(x,t)$. \\ \\
This is a very surprising analogy. To understand it deeply, we will take a physical point of view of the dynamical optimization problem. We will consider it as a physical dynamic system and explore its characteristics from both the classical and quantum mechanics points of view. We hope to found the relations between Pontryagin's scheme and Bellman's dynamic programming, from a physicist's perspective. We start with the classical vision. \\ \\
Consider again the Lagrangian (\ref{Lagrangiancontopt2}). If we want to think of $x$ as a position variable for a certain physical particle and of $p_x$ as its canonical momentum, the Lagrangian (\ref{Lagrangiancontopt2}) has the wrong sign. To consider this system as a physical one, we would take instead the Lagrangian
\begin{equation} \label{Lagrangiancontopt3}
	L'(x,u, \dot{x}, \dot{u}) = p_x \dot{x} - ( F(x,u,t) + p_x f(x,u,t)).
\end{equation}
Obviously, these Lagrangians have the same equations of motion. Now, we analyze (\ref{Lagrangiancontopt3}) from the  phase space $(x, u, p_x, p_u,)$ point of view; that is, we formulate a Hamiltonian theory related to (\ref{Lagrangiancontopt3}). To do that, we first must note that, in the Lagrangian (\ref{Lagrangiancontopt3}), the momentum definition for the variable $u$ is
\begin{eqnarray*}
	p_u = \frac{\partial L'(x,u, \dot{x}, \dot{u},)}{\partial \dot{u}} =  0 .
\end{eqnarray*}
Note that in this case, the momentum variable definition does not allows us to write $p_u$ in terms of its respective velocity $\dot{u}$, so the momentum definition gives origin to one constraint (this is the same problem that appeared in the Lagrangian (\ref{constraint1})). To this point, we need to apply Dirac's method to study the system in the right way. According to Dirac's classification, the above constraint is called primary constraint, and we write it as:
\begin{eqnarray} \label{Phi-1}
	\Phi_1 = p_u = 0 .
\end{eqnarray}
The Hamiltonian $H=H(x, u, p_x, p_u)$ is
\begin{equation*}
	H = p_x \dot{x} + p_u \dot{u} -  L'(x,u, \dot{x}, \dot{u}),
\end{equation*}
which expands into
\begin{equation*}
	H = p_x \dot{x} + p_u \dot{u} + ( F(x,u,t) + p_x f(x,u,t)- p_x \dot{x} ) ,
\end{equation*}
or
\begin{equation*}
	H= \Phi_1 \dot{u} + H_0(x, u, p_x, p_u) ,
\end{equation*}
with
\begin{equation} \label{Hcero}
	H_0(x, u, p_x, p_u) = F(x,u,t) + p_x f(x,u,t).
\end{equation}
For the constraint surface $\Phi_1 = 0$, we get
\begin{equation*}
	H(x, u, p_x, p_u)=H_0(x, u, p_x, p_u) .
\end{equation*}
To incorporate these restrictions over the phase space, we define the extended Hamiltonian
\begin{equation} \label{Hextended}
	\tilde{H}(x, u, p_x, p_u) = H_0 + \mu_1 \Phi_1 ,
\end{equation}
where $\mu_1$ is a Lagrange multiplier. Now, we require the constraint $\Phi_1$ to be preserved in time with the extended Hamiltonian (\ref{Hextended}) such that
\begin{eqnarray*} \label{dotPhi1}
	\dot{\Phi}_1= \{\Phi_1, \tilde{H} \} = 0 .
\end{eqnarray*}
This equation gives,
\begin{equation} \label{dotPhi2v2}
	\dot{\Phi}_1= \{p_u, \tilde{H} \} =  \frac{\partial \tilde{H}}{\partial u} =  \frac{\partial H_0}{\partial u} = 0 .
\end{equation}
Thus, (\ref{dotPhi2v2}) is a new secondary constraint:
\begin{equation} \label{Phi-4}
	\Phi_2 = \frac{\partial H_0}{\partial u} = \frac{\partial F(x,u,t)}{\partial u} + p_x \frac{\partial f(x,u,t)}{\partial u} = 0.
\end{equation}
In this way, the optimization of the Hamiltonian with respect to the control variable appears in the phase space as a secondary constraint. To incorporate the new constraint in the model, we must consider the Hamiltonian
\begin{equation*} \label{Hextended2}
	\tilde{H}_2(x, u, p_x, p_u) = H_0 + \mu_1 \Phi_1 + \mu_2 \Phi_2 .
\end{equation*}
We start again and impose time preservation for the constraints set $\{ \Phi_1, \Phi_2 \}$ with the new Hamiltonian  $ \tilde{H}_2 $:
\begin{eqnarray*} \label{dotPhi1}
	\dot{\Phi}_1= \{\Phi_1, \tilde{H}_2 \} = 0 ,\\
	\dot{\Phi}_2= \{\Phi_2, \tilde{H}_2 \} = 0 .
\end{eqnarray*}
The above set of two equations gives only restrictions for the Lagrange multipliers $\mu_1, \ \mu_2$ and no new constraint appears. In fact these equations are explicitly
\begin{eqnarray}
	\{\Phi_1, H_0 \} + \mu_2 \{\Phi_1, \Phi_2 \} = 0   \label{dotPhi1v2_1} \\
	\{\Phi_2, H_0 \} + \mu_1 \{\Phi_2, \Phi_1 \} = 0 ,  \label{dotPhi1v2_2}
\end{eqnarray}
or in matrix form
\begin{equation*}
	\left( \begin{array}{c}   \{\Phi_1, H_0 \}  \\
		\{\Phi_2, H_0 \}
	\end{array} \right) +  \left( \begin{array}{c c}  0 &  \{\Phi_1, \Phi_2 \}  \\
	- \{\Phi_1, \Phi_2 \}   &  0
\end{array} \right)
\left( \begin{array}{c}  \mu_1  \\
	\mu_2
\end{array} \right) = \left( \begin{array}{c}  0  \\
0
\end{array} \right) .
\end{equation*}
The antisymmetric matrix
\begin{equation*}
	\Delta =  \left( \begin{array}{c c}  0 &  \{\Phi_1, \Phi_2 \}  \\
		- \{\Phi_1, \Phi_2 \}   &  0
	\end{array} \right) ,
\end{equation*}
is called the Dirac matrix. Now
\begin{equation*}
	\{\Phi_1, \Phi_2 \} = \{p_u, \Phi_2 \} =\frac{\partial \Phi_2}{\partial u} =  \frac{\partial^2 H_0}{\partial u^2}  ,
\end{equation*}
so
\begin{equation*}
	\Delta = \left( \begin{array}{c c}
		0 & \frac{\partial^2 H_0}{\partial u^2} \\
		- \frac{\partial^2 H_0}{\partial u^2} &  0 \\
	\end{array} \right) .
\end{equation*}
The determinant of the Dirac matrix is
\begin{equation} \label{detDM}
	\det(\Delta) =   \Big( \frac{\partial^2 H_0}{\partial u^2} \Big)^2.
\end{equation}
If
\begin{equation} \label{SCcond}
	\frac{\partial^2 H_0}{\partial u^2} \ne 0 ,
\end{equation}
on the constraint surface where $\Phi_1=0, \Phi_2=0$, then the Dirac matrix is invertible and the constraint set $ \{ \Phi_1, \ \Phi_2 \} $ is second-class (see \cite{dirac1}, \cite{dirac2}, \cite{contrerasbustamante}). \\ \\
For the rest of the paper, we will assume that (\ref{SCcond}) is valid (for example, in a typical control problem in economics, the function $F$ is quadratic and the function $f$ is linear in terms of the control variable, so $\frac{\partial^2 H_0}{\partial u^2} \ne 0$). Thus, the optimization problem defined by the Lagrangian (\ref{Lagrangiancontopt3}), from a physical point of view, corresponds to a second-class constrained dynamic system in the phase space. \\ \\
Now equations (\ref{dotPhi1v2_1}), (\ref{dotPhi1v2_2}) can be used to obtain the Lagrange multipliers $\mu_1$ and $\mu_2$ as
\begin{equation*}
	\left( \begin{array}{c}  \mu_1  \\
		\mu_2
	\end{array} \right) = - \Delta^{-1}
	\left( \begin{array}{c}   \{\Phi_1, H_0 \}  \\
		\{\Phi_2, H_0 \}
	\end{array} \right) ,
\end{equation*}
that is
\begin{equation} \label{lagranmultipliers}
	\left( \begin{array}{c}  \mu_1  \\
		\mu_2
	\end{array} \right) = \left( \begin{array}{c}  \frac{1}{\{\Phi_1, \Phi_2 \}} \{\Phi_2, H_0 \}  \\
	-\frac{1}{\{\Phi_1, \Phi_2 \}} \{\Phi_1, H_0 \}
\end{array} \right) .
\end{equation}
The time evolution of any function $F$ over the phase space generated by the Hamiltonian $\tilde{H}_2$ is
\begin{equation*}
	\dot{F} = \{ F , \tilde{H}_2 \} =  \{F, H_0 \} + \mu_1  \{ F, \Phi_1 \} + \mu_2 \{ F, \Phi_2 \} .
\end{equation*}
Substituting the Lagrangian multipliers (\ref{lagranmultipliers}) in the above equation we obtain
\begin{equation} \label{diracevol}
	\begin{array}{c l}  \dot{F} = & \{F, H_0 \} +
		\{ F, \Phi_1 \} \frac{1}{\{\Phi_1, \Phi_2 \}} \{\Phi_2, H_0 \} \ +   \\
		& \\
		& \{ F, \Phi_2 \} \frac{1}{\{\Phi_2, \Phi_1 \}} \{\Phi_1, H_0 \} .
	\end{array}
\end{equation}
The whole expression in the right hand side of (\ref{diracevol}) is called the Dirac bracket, defined by
\begin{equation*}
	\begin{array}{c l} \{  A , B  \}_{DB} = &\{ A , B \} +  \{ A, \Phi_1 \}\Delta^{-1}_{1 2} \{\Phi_2, B \} +    \\
		& \\
		& \{ A, \Phi_2 \}  \Delta^{-1}_{2 1} \{\Phi_1, B \} ,
	\end{array}
\end{equation*}
so we can write (\ref{diracevol}) as
\begin{equation*}
	\dot{F} = \{F, H_0 \}_{DB} .
\end{equation*}
The dynamical evolution of the variables $x, p_x, u, p_u$ in the phase space, in the presence of the second-class constraints $\Phi_1$, $\Phi_2$  is then, given by
\begin{equation*}
	\begin{array}{rcl}
		\dot{x} & = & \{x, H_0 \}_{DB}  \\
		\dot{p}_x & = & \{p_x, H_0 \}_{DB}   \\
		\dot{u} & = & \{u, H_0 \}_{DB}  \\
		\dot{p_u} & = & \{p_u, H_0 \}_{DB} .
	\end{array}
\end{equation*}
The last equation is
\begin{equation}
	\dot{p_u} = \{\Phi_1, H_0 \}_{DB} =  \{\Phi_1, \tilde{H}_2 \} = 0 ,
\end{equation}
which is consistent with the time preservation of $\Phi_1 = 0$.\\
The Dirac bracket between the second-class constraints $\Phi_1, \Phi_2$ \ is
\begin{equation*}
	\begin{array}{r l} \{\Phi_1, \Phi_2 \}_{DB} = & \{ \Phi_1 , \Phi_2 \} +
		\{ \Phi_1, \Phi_1 \} \frac{1}{\{\Phi_1, \Phi_2 \}} \{\Phi_2, \Phi_2 \} \ \\
		& \\
		+ & \{ \Phi_1, \Phi_2 \} \frac{1}{\{\Phi_2, \Phi_1 \}} \{\Phi_1, \Phi_2 \} \\
		& \\
		= & 0 . \\
	\end{array}
\end{equation*}
Thus, the use of the Dirac bracket is equivalent to eliminate the second-class constraints from the theory or, which is the same, to set all second constraints to zero. \\ \\
Now we compute explicitly the dynamic behavior of variables $x$ and $p_x$ for the constrained system, that is
\begin{equation*}
	\begin{array}{rcl}  \dot{x} & = & \{x, H_0 \}_{DB} \\
		&&   \\
		& = & \{x, H_0 \} + \{ x, \Phi_1 \} \frac{1}{\{\Phi_1, \Phi_2 \}} \{\Phi_2, H_0 \}   \\
		&&   \\
		& + & \{ x, \Phi_2 \} \frac{1}{\{\Phi_2, \Phi_1 \}} \{\Phi_1, H_0 \} \\
		&&   \\
		& = & \frac{\partial H_0}{\partial p_x} + \{ x, p_u \} \frac{1}{\{\Phi_1, \Phi_2 \}} \{\Phi_2, H_0 \}   \\
		&&   \\
		& + &\{ x, \Phi_2 \} \frac{1}{\{\Phi_2, \Phi_1 \}} \{ p_u, H_0 \} ,
	\end{array}
\end{equation*}
but $\{ x, p_u \} = 0 $ and $\{ p_u, H_0 \} = - \frac{\partial H_0}{\partial u} =-\Phi_2 = 0$, so
\begin{equation*}
	\dot{x} = \{x, H_0 \}_{DB}  =  \frac{\partial H_0}{\partial p_x} .
\end{equation*}
In the same way we obtain, for the momentum,
\begin{equation*}
	\begin{array}{rcl}  \dot{p_x} & = & \{p_x, H_0 \}_{DB}\\
		& &  \\
		& = & \{p_x, H_0 \} + \{p_x, \Phi_1 \} \frac{1}{\{\Phi_1, \Phi_2 \}} \{\Phi_2, H_0 \} \\
		& &  \\
		& + & \{p_x, \Phi_2 \} \frac{1}{\{\Phi_2, \Phi_1 \}} \{\Phi_1, H_0 \} \\
		& &  \\
		& = & - \frac{\partial H_0}{\partial x} + \{ p_x, p_u \} \frac{1}{\{\Phi_1, \Phi_2 \}} \{\Phi_2, H_0 \}\\
		& & \\
		& + & \{ p_x, \Phi_2 \} \frac{1}{\{\Phi_2, \Phi_1 \}} \{ p_u, H_0 \} ,
	\end{array}
\end{equation*}
but $\{ p_x, p_u \} = 0 $, then
\begin{equation*}
	\dot{p_x} = \{p_x, H_0 \}_{DB}  = - \frac{\partial H_0}{\partial x} .
\end{equation*}
Thus, the constrained dynamic given by the Dirac bracket is the same unconstrained dynamic given by the Pontryagin equations. These Pontryagin equations are the classical equations of motion for our physical constrained system.

\section{Dynamic optimization and quantum mechanics}

Until now, we have studied the dynamic optimization problem from a classical point of view, and we have seen that it is equivalent to a classical physically constrained system. However, what happens at the quantum level? To explore that view, we will quantize our classical system and study its consequences. \\ \\
Again, consider the classical Hamiltonian
\begin{equation} \label{QH1}
	H_0(x, u, \lambda, p_x, p_u) =  F(x,u,t) + f(x,u,t) \ p_x .
\end{equation}
Now, we have to have to quantize the classical Hamiltonian (\ref{QH1}). For this purpose, we replace $p_x, p_u$ with the appropriate quantum momentum operators (in the usual sense, with $\hbar = 1$):
\begin{eqnarray*}
	\hat{p}_x = - i \frac{\partial}{\partial x}   \\
	\hat{p}_u = - i \frac{\partial}{\partial u} ,
\end{eqnarray*}
thus, the Schr\"{o}dinger equation
\begin{equation*}
	\hat{H}_0(x, u, \hat{p}_x, \hat{p}_u) \ \Psi(x, u, t)  = i \frac{\partial }{\partial t} \ \Psi (x, u, t)
\end{equation*}
writes as:
\begin{equation} \label{SReq1}
	\Bigl( F(x,u,t) - i f(x,u,t) \frac{\partial}{\partial x} \Bigr)  \\ \Psi  = i \frac{\partial \Psi}{\partial t} \ ,
\end{equation}
where $\Psi = \Psi (x, u, t)$ (note that we have choosen right-hand side order for the momentum operator in the quantization process). \\ \\
According to Dirac, not all solutions $\Psi$ in the Schr\"{o}dinger equation (\ref{SReq1}) are physically admissible. The physical solutions $\Psi_P$ must satisfy the constraint equations (\ref{Phi-1}) and (\ref{Phi-4}). In the quantum case, these equations must be imposed as operator equations over the Hilbert space of states in the form
\begin{eqnarray} \label{QPhi1}
	\nonumber\hat{\Phi}_1 \ \Psi_P = 0 ,\\
	\hat{\Phi}_2 \ \Psi_P = 0 .
\end{eqnarray}
The physically admissible states are the solutions of (\ref{SReq1}) that satisfy the constraints in equations (\ref{QPhi1}). Explicitly, we can define these constraints operators as
\begin{eqnarray*} \label{QQPhi1}
	\hat{\Phi}_1 & = & \hat{p}_u = - i \frac{\partial }{\partial u}\\
	\hat{\Phi}_2 & = & \frac{\partial F(x,u,t) }{\partial u} - i \frac{\partial f(x,u,t) }{\partial u} \frac{\partial }{\partial x} .
\end{eqnarray*}
Thus, physically admissible solutions $\Psi_P (x, u, t) $, must satisfy
\begin{equation} \label{QQQPhi2}
	- i \frac{\partial }{\partial u} \Psi_P  = 0,
\end{equation}\vspace{-0.5cm}
\begin{equation}\label{QQQPhi4}
	\Bigg( \frac{\partial F(x,u,t) }{\partial u} - i \frac{\partial f(x,u,t) }{\partial u} \frac{\partial }{\partial x} \Bigg) \Psi_P  = 0,
\end{equation}
and the Schr\"{o}dinger equation
\begin{equation}\label{QQQPhi5}
	\Bigl( F(x,u,t) - i f(x,u,t)  \frac{\partial }{\partial x} \Bigr) \ \Psi_P   = i \frac{\partial }{\partial t} \ \Psi_P .
\end{equation}
Equation (\ref{QQQPhi2}) implies that the wave function is independent of $u$, thus $\Psi_P  =  \Psi_P (x,t)$. Now, if we define the function $S(x,t)$ as
\begin{equation} \label{Sdext}
	S(x,t)= \ln( - i \ \Psi_P (x,t) ) ,
\end{equation}
the quantum Schr\"{o}dinger's equations (\ref{QQQPhi4}) and (\ref{QQQPhi4}) can be writing in terms of $S(x,t)$ as
\begin{equation} \label{QQ1}
	\frac{\partial F(x,u,t) }{\partial u} \  + \frac{\partial f(x,u,t) }{\partial u} \frac{\partial \  S(x,t)}{\partial x} = 0,
\end{equation}
\begin{equation} \label{QQ2}
	F(x,u,t) + f(x,u,t) \frac{\partial S(x,t) }{\partial x} = - \frac{\partial S(x,t) }{\partial t}.
\end{equation}
The constraint (\ref{QQ1}) is the expression for the maximization of the quantity
\begin{equation*}
	F(x,u,t) + f(x,u,t) \frac{\partial S(x,t) }{\partial x} ,
\end{equation*}
with respect to the control variable $u$, so equations (\ref{QQ1}) and (\ref{QQ2}) can be written as a unique equation:
\begin{equation} \label{QHJB}
	\max_u \Bigl( F(x,u,t) + f(x,u,t) \frac{\partial S(x,t) }{\partial x} \Bigr) = - \frac{\partial S(x,t) }{\partial t},
\end{equation}
which is the Hamilton-Jacobi-Bellman equation (\ref{HJB}) if the $S(x,t)$ function is identified with the optimal value function $J(x,t)$. Thus, the dynamic programming approach to the dynamic optimization problem corresponds to a quantum view. \\ \\
The Dirac's quantization method used have some problems. The commutator of the constraint operators $\hat{\Phi}_1$ and $\hat{\Phi}_2$ is, in general, different from zero, thus
\begin{equation} \label{qcmt}
	[\hat{\Phi}_1, \hat{\Phi}_2] = \alpha(x,u) \hat{I} \ne 0 ,
\end{equation}
for some function $\alpha$, in such a way that, when applied to a physical state $\Psi_P$, we find
\begin{equation*}
	[\hat{\Phi}_1, \hat{\Phi}_2] \ \Psi_P = \alpha(x,u) \Psi_P .
\end{equation*}
This implies that $0 = \alpha(x,u) \Psi_P$ and $0 = \Psi_P$. Thus, there is not physical wave function at all. \\ \\
This is related to the fact that the Poisson bracket of two second-class constraints $\{ \Phi_1, \Phi_2 \}$ is different from zero.  After the quantization, the Poisson bracket becomes the commutator (\ref{qcmt}). Then, Dirac's quantization procedure is not well-defined for second-class constraints (for details related to this issue see \cite{Klauder}). \\ \\
Note that we have quantized the model without solving the classical constraints. A more transparent procedure would be to solve the classical constraints first, and then, to quantize (these procedures give, in general, different answers; see \cite{Klauder}). In this way, we can solve the classical constraint (\ref{Phi-1}) first simply by setting $p_u = 0$. Let
\begin{equation} \label{upxt}
	u = u^{*}(x,p_x,t)
\end{equation}
be the solution of the constraint (\ref{Phi-4}) for $u$ in terms of $x$ and $p_x$. These solutions can be substituted in the classical Hamiltonian (\ref{Hcero}), in such a way that we end with a reduced Hamiltonian $H^{*}_0$ that depends only on $x$ and $p_x$:
\begin{equation} \label{H0xpxt}
	\begin{array}{ll}
		H^{*}_0 (x,p_x,t) & = H_0 (x, p_x, u^{*}(x,p_x,t),t) ,  \\
		&   \\
		& = F(x,u^{*},t) + p_x f(x,u^{*}, t).
	\end{array}
\end{equation}
Quantizing the above Hamiltonian by replacing
\begin{equation*}
	p_x \rightarrow \hat{p}_x = - i \frac{\partial}{\partial x} ,
\end{equation*}
the Schr\"{o}dinger equation reads in this case
\begin{equation}\label{sxt}
	F(x,u^{*}(x,\hat{p}_x,t),t) + \hat{p}_x f(x,u^{*}(x,\hat{p}_x,t)) \Psi = i \frac{\partial \Psi }{\partial t} .
\end{equation}
By replacing the relation (\ref{Sdext}) in (\ref{sxt}), we obtain a non-linear differential equation for $S(x,t)$. Note that, when right-hand side order is taken in the quantization procedure, the substitution of (\ref{Sdext}) in the  Schr\"{o}dinger equation (\ref{sxt}) ends with the same Hamilton-Jacobi-Bellman equation (\ref{HJBv3}). Thus, the Hamilton-Jacobi-Bellman equation of the dynamic programming method correspond to the right quantization procedure of the our classical constrained system associated to the dynamical optimization problem.

\section{Conclusions}
In this article, we have examined the structure of the dynamic optimization problem from a physical perspective, and we conclude that the correct analysis of the optimization problem must be done either in the phase-space or using the classical Hamiltonian approach. Due to the presence of constraints in the theory, we must apply Dirac's method for constrained systems. Dirac's analysis implies that the theory has two second-class constraints. One of these constraints fixes the momentum associated to the control variable,  and the other represents the optimization of the Hamiltonian respect to the control variable. \\

The dynamic evolution of this constrained system is given by Dirac's brackets of the canonical variables with the Hamiltonian. This dynamic results to be identical to the unconstrained one given by the Pontryagin equations. Thus, the Pontryagin equations are the correct classical equations of motion of our physical optimization problem. \\

In the same Pontryagin scheme, by imposing a closed-loop $\lambda$-strategy, the optimality of the action gives a consistency relation for $\lambda(x,t)$. This consistency relation is connected to the Hamilton-Jacobi-Bellman equation. The solution of the optimization problem in this $\lambda$-closed-loop case is obtained by the Pontryagin equation for the $x$ coordinate, evaluated over the optimal $\lambda^{*}$-strategy that satisfies the consistency condition. \\

The same result is achieved by quantizing the classical constrained model. By writing the wave function in the form $\Psi(x,t) = e^{iS(x,t)}$ and substituting it into the quantum Schr\"{o}dinger equation, a non-linear partial differential equation is obtained for the $S$ function. For the right-hand side quantization, this non-linear equation is just the Hamilton-Jacobi-Bellman equation when $S(x,t)$ is identified with the optimal value function $J(x,t)$. \\ \\
Thus, we end this paper by concluding that Bellman's maximum principle of the dynamic programming method corresponds to the right-hand side quantization of the Pontryagin theory.


\begin{thebibliography}{25}
	
	\bibitem{Cbennatirosaclot} E. Bennati, M. Rosa-Clot and S. Taddei, \textit{A path integral approach to derivative security pricing I}, Int. J. Theor. Appl. Fin., \textbf{2}, 4, 381, (1999).
	
	\bibitem{Otto} M. Otto, \textit{Using path integrals to price interest rate derivatives}, cond-mat/9812318v2, (1999).
	
	\bibitem{lemmens1} D. Lemmens, M. Wouters and J. Tempere, \textit{A path integral approach to closed-form option pricing formulas with applications to stochastic volatility and interest rate models}, q-fin.PR arXiv 0806.0932v1, (2008).
	
	\bibitem{baaquie} B. E. Baaquie, \textit{Quantum Finance: Path Integrals and Hamiltonians for Option and Interest Rates}, (Cambridge, 2007).
	
	\bibitem{baaquie2} B. E. Baaquie, \textit{A Path Integral to Option Price with Stochastic Volatility: Some Exact Results}, J. Phys. I France, \textbf{7}, 12, 1733-1753, (1997).
	
	\bibitem{Linetsky} V. Linetsky, \textit{The path integral approach to financial modelling and option pricing}, Comput. Econ. \textbf{11}, 129--163, (1998).
	
	\bibitem{Dash} Jan W. Dash, \textit{Quantitative Finance and Risk Management: A Physicist Approach}, (World Scientific, 2004).
	
	\bibitem{Haven1} E. Haven, \textit{A Black-Scholes Schrödinger option price: "bit" versus "qubit"}, Physica A, (2003).
	
	\bibitem{Haven2} E. Haven, \textit{A discussion on embedding the Black-Scholes option price model in a quantum physics setting}, Physica A, (2002).
	
	\bibitem{semiclass} M. Contreras, R. Pellicer, M. Villena and A. Ruiz \textit{A quantum model for option pricing: when Black-Scholes meets Schrödinger and its semi-classic limit}, Physica A, \textbf{329} (23) 5447-5459, (2010).
	
	\bibitem{dirac1} Paul M. Dirac, \textit{Generalized Hamiltonian Dynamics}, in Proc. Roy. Soc., London, \textbf{A246} 326, (1958).
	
	\bibitem{dirac2} Paul M. Dirac, \textit{Lectures on quantum mechanics}, Yeshiva University, (1967).
	
	\bibitem{contrerashojman} M. Contreras and S. Hojman \textit{Option pricing, stochastic volatility, singular dynamics and constrained path integrals}, Physica A, \textbf{393}, 1, 391-403, (2014).
	
	\bibitem{contreras} M. Contreras \textit{Stochastic volatility models at $\rho = \pm 1$  as a second-class constrained hamiltonian systems}, Physica A, \textbf{405} 289-302, (2015).
	
	\bibitem{contrerasbustamante} M. Contreras and M. Bustamante \textit{Multi-asset Black-Scholes model as a variable second-class constrained dynamical system}, Physica A, \textbf{457}, 540-572, (2016).
	
	\bibitem{erikson1} G. M. Erickson, \textit{Differential  game models of advertising competitions}, J. of Political Economy \textbf{8}, 31, 637-654, (1973).
	
	\bibitem{Bellman}  R. Bellman \textit{The theory of dynamic programming}, Bull. Am. Math. Soc. \textbf{60}, 6, 503-516, (1954).
	
	\bibitem{Kamien}  M. I. Kamien and N. L. Schwartz \textit{Dynamic Optimization: The Calculus of Variations and Optimal Control in Economics and Management}, (Advanced Textbooks in Economics, Elsevier Science, 2nd Ed., 1991).
	
	\bibitem{Klauder} J. R. Klauder, \textit{A modern approach to functional integration}, (Birkh\"{a}user, 2010).
	
	
\end{thebibliography}
\end{document}